\documentclass[12pt]{iopart}

\usepackage{graphicx}

\begin{document}


\title{Extrapolation of Multiplicity distribution in $p$+$p$($\bar{p}$) collisions to LHC energies}
\author{Ajay Kumar Dash$^{1}$ and Bedangadas Mohanty$^{2}$}
\address{$^{1}$ Institute Of Physics, Bhubaneswar 751005, India 
$^{2}$ Variable Energy Cyclotron Centre, Kolkata 700064, India}
\ead{ajay@iopb.res.in,bmohanty@veccal.ernet.in}

\begin{abstract}
The multiplicity ($N_{\mathrm {ch}}$) and pseudorapidity distribution ($dN_{\mathrm {ch}}/d\eta$)
of primary charged particles
in $p$+$p$ collisions at Large Hadron Collider (LHC) energies of
$\sqrt{s}$ = 10 and 14 TeV are obtained from extrapolation of existing
measurements at lower $\sqrt{s}$. These distributions are then compared
to calculations from PYTHIA and PHOJET models. The existing $\sqrt{s}$
measurements are unable to distinguish between a logarithmic and power law
dependence of the average charged particle multiplicity ($\langle N_{\mathrm {ch}} \rangle$)
on $\sqrt{s}$, and their extrapolation to energies accessible at LHC give very different values.
Assuming a reasonably good description of inclusive charged particle multiplicity
distributions by Negative Binomial Distributions (NBD) at lower $\sqrt{s}$ to
hold for LHC energies, we observe that the logarithmic $\sqrt{s}$ dependence of
$\langle N_{\mathrm {ch}} \rangle$ are favored by the models at midrapidity.
The $dN_{\mathrm {ch}}/d\eta$ versus $\eta$ for the existing measurements are
found to be reasonably well described by a function with three parameters
which accounts for the basic features of the distribution, height at midrapidity, central rapidity
plateau and the higher rapidity fall-off. Extrapolation of these
parameters as a function of $\sqrt{s}$ is used to predict the pseudorapidity distributions of
charged particles at LHC energies. $dN_{\mathrm {ch}}/d\eta$ calculations from PYTHIA and PHOJET models
are found to be lower compared to those obtained from the extrapolated
$dN_{\mathrm {ch}}/d\eta$ versus $\eta$ distributions for a broad $\eta$
range.

%
\end{abstract}
\maketitle
\section{Introduction}
\label{intro}

The Large Hadron Collider (LHC) at CERN is designed
for colliding proton-proton beams upto $\sqrt{s}$ = 14 TeV~\cite{lhc}.
Collisions at these unprecedented high energies will
provide opportunities for new physics~\cite{lhc}. In order to fully
exploit the enormous physics potential it is important
to have a complete understanding of the reaction mechanism.
The particle multiplicity distributions, one of the first
measurements to be made at LHC, will be used to test various
particle production models based on different physics
mechanism and also provide constrains on model features.
Some of these models are based on  string fragmentation
mechanism~\cite{pythia} and some are based on Pomeron exchange~\cite{phojet}.

In this paper, we first make a compilation of the existing
data on the average charged particle multiplicity (at midrapidity
and full rapidity range) and charged particle pseudorapidity distribution
as a function of $\sqrt{s}$. Then we judiciously extrapolate the measurements
to obtain prediction of $\langle N_{\mathrm {ch}} \rangle$ (for
midrapidity and full rapidity range), charged particle multiplicity
distributions and $dN_{\mathrm {ch}}/d\eta$ distributions
at LHC energies of $\sqrt{s}$ = 10 and 14 TeV. These results are
also compared to calculations from PYTHIA \cite{pythia} 
and PHOJET \cite{phojet} monte carlo models.
Energy dependence of average number of particles produced in 
$p$+$p$($\bar{p}$) collisions 
can be used to distinguish
various models~\cite{malhotra}. A statistical model~\cite{fermi} and
some hydrodynamical models~\cite{landau}
predict a dependence of $\langle N \rangle$ $\sim$ $s^{1/4}$, whereas
multiperipheral models~\cite{detar} and Feynman's scaling~\cite{feynman}
lead to the dependence
as $\langle N \rangle$ $\sim$ $ln(s)$. Such a logarithmic dependence and 
on higher powers of $ln(s)$ is also predicted by Regge-Mueller model~\cite{regge}. 
Arguments based on simple
phase space considerations~\cite{satz} however predict a power law dependence
as $\langle N \rangle$ $\sim$ $s^{1/3}$. In this work we show
that for the $\langle N_{\mathrm {ch}} \rangle$ data available at various $\sqrt{s}$,
we cannot distinguish the logarithmic and power law dependences on $\sqrt{s}$.
However the measurements at the LHC energies will provide a clear answer. The
pseudorapidity distributions on the other hand is found to be described by a 
form which resembles a generalized Fermi distribution. Such distributions 
have been used to explain the pseudorapidity distributions of produced particles
in hadronic collisions at ISR~\cite{isr} and heavy-ion collisions at RHIC~\cite{pmd}.

The compiled experimental data presented in this paper corresponds to Non Singly Diffractive
(NSD) events for minimum bias triggers. The charged particle data were corrected for
secondary interactions, gamma conversions, short lived decays ($K^{0}_{S}$, $\Lambda$),
reconstruction efficiency and acceptance effects by the experiments. To match the
experimental conditions, the model simulations presented are also corrected for
short lived decays. A transverse momentum ($p_{T}$) cut of greater than 100 MeV/$c$
are usually used in realistic experimental conditions, as for example in ALICE
experiment at LHC. Model simulations using PYTHIA and PHOJET 
suggest, a 6\% $\pm$ 2\% effect on the charged hadron
multiplicity due to a 100 MeV/$c$ cut-off in $p_{T}$ at $\sqrt{s}$ = 10-14 TeV. The error
of 2\% comes from the difference in results from PYTHIA and PHOJET models. 
At LHC, while most of the experiments will have mid-rapidity
measurements of charged particle multiplicity, ALICE experiment has the possibility
to measure the distributions over -5.0 $<$ $\eta$ $<$ 3.5 range~\cite{lhc}. The CMS and ATLAS
experiments will have a more limited coverage of $|\eta|$ $<$ 2.5 units~\cite{lhc}.

The model results presented are from PYTHIA using
version 6.4 (ATLAS tuned) and those from PHOJET with version 1.12 (default settings). 
It may be mentioned that recently a new tuned version of 
PYTHIA has been released~\cite{pythiatuned}.
The PHOJET model combines the ideas based on a dual parton model~\cite{dpm} on
soft process of particle production and uses lowest-order perturbative QCD for
hard process. Regge phenomenology is used to parameterize the total, elastic and
inelastic cross-sections. The initial and final state parton shower are generated in
leading log-approximation. PYTHIA on the other hand uses string fragmentation as a
process of hadronization and tends to use the perturbative parton-parton scattering
for low to high $p_{T}$ particle production.
Although there are several other theoretical predictions on total cross section
expected at LHC energies~\cite{cross}, current work focuses on how a judicious extrapolation
from existing multiplicity data compares to the calculations from some of the available models~\cite{pythia,phojet}.

\section{Multiplicity distribution}
\label{sec:0}

\begin{figure}[ht]
\begin{center}
 \includegraphics[scale=0.35]{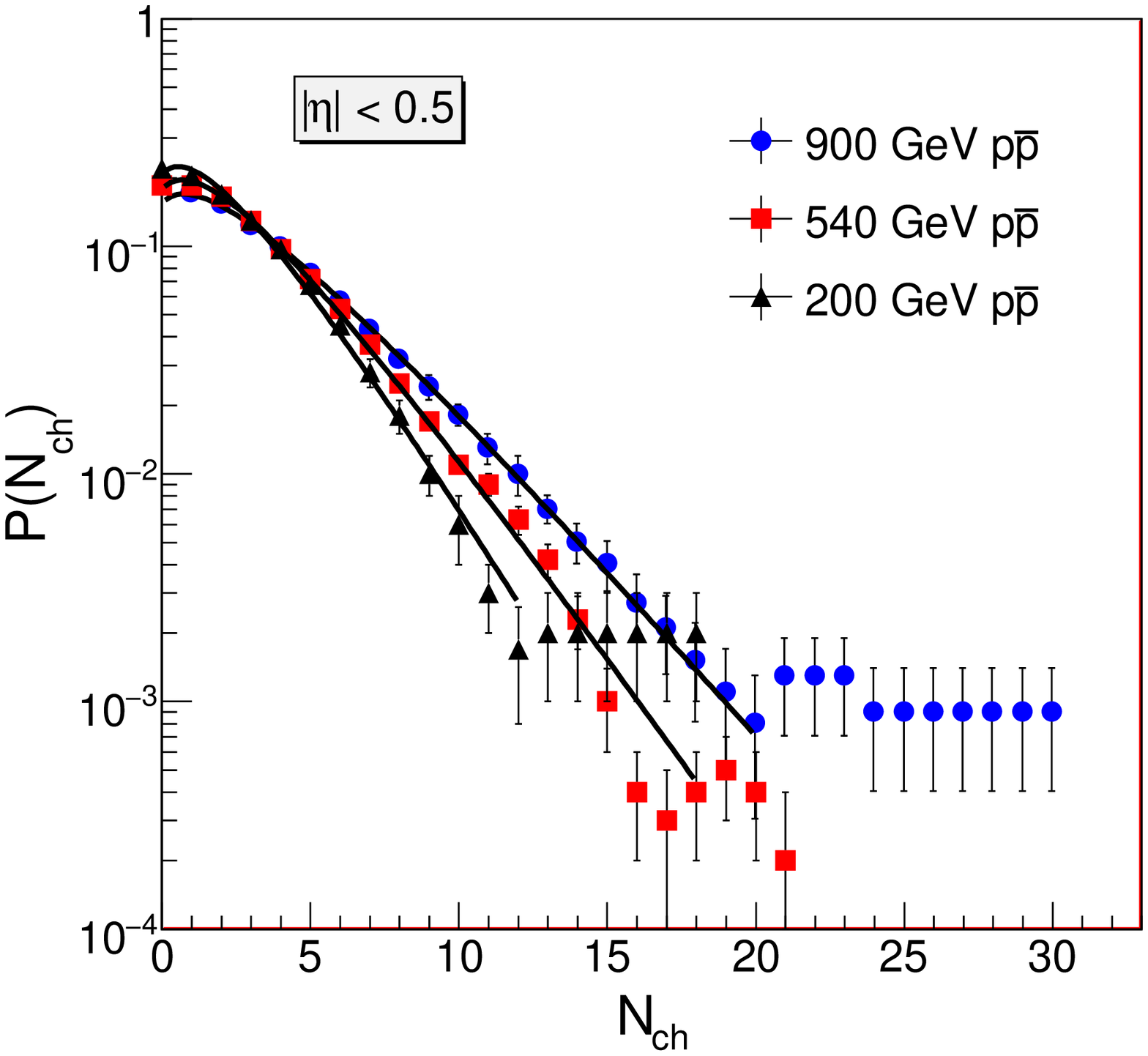}
 \includegraphics[scale=0.35]{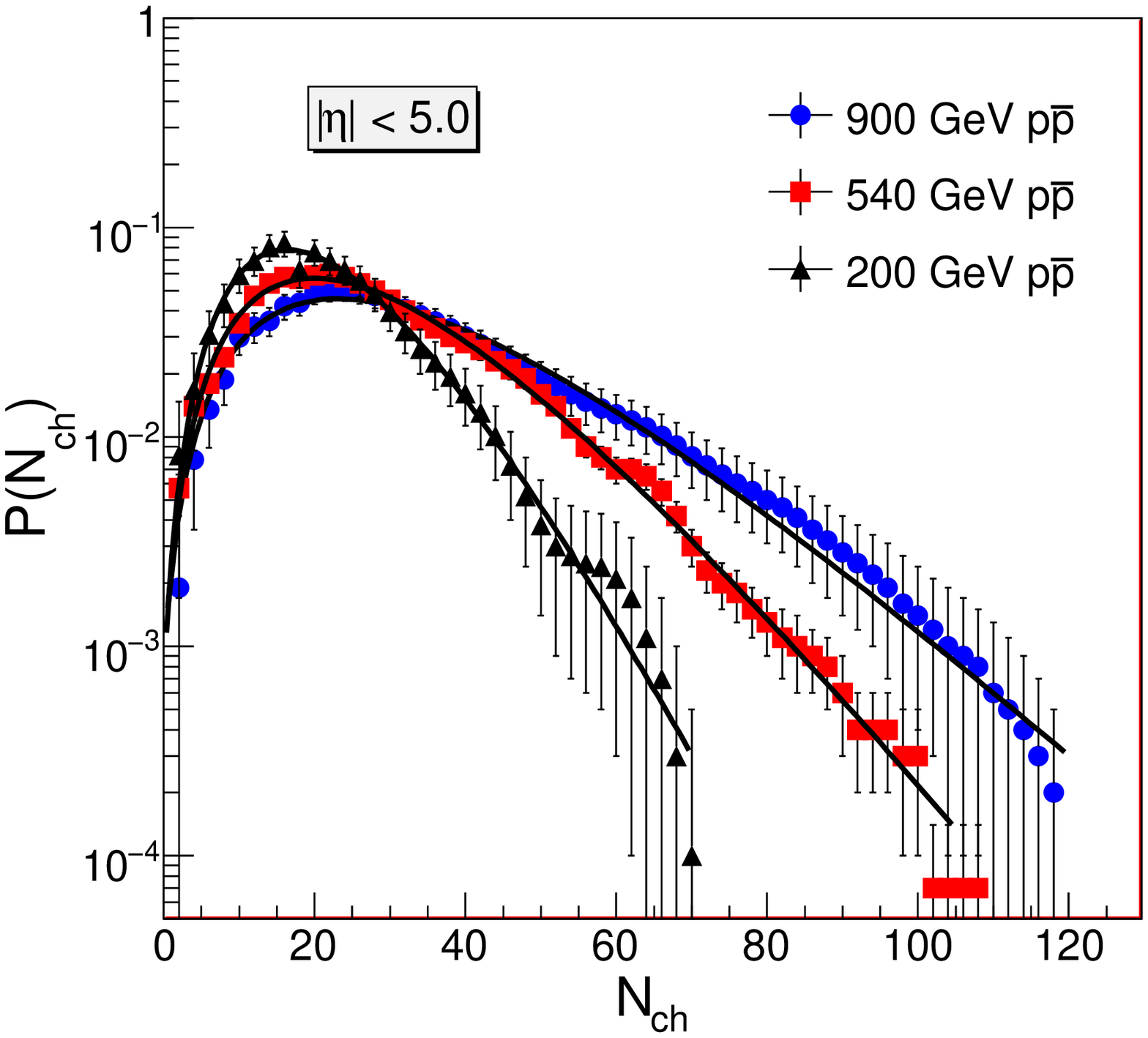}
\caption{Multiplicity distribution for charged particles in
$p$+$\bar{p}$ collisions at various center of mass energies at midrapidity (left panel) and
full rapidity (right panel) ranges~\cite{datamult}. The errors are statistical.
The solid lines are NBD fit to the data points using the function given in
Eqn.~\ref{nbdfit}.  }
\label{nbd}       
\end{center}
\end{figure}

\begin{figure}
\begin{center}
 \includegraphics[scale=0.35]{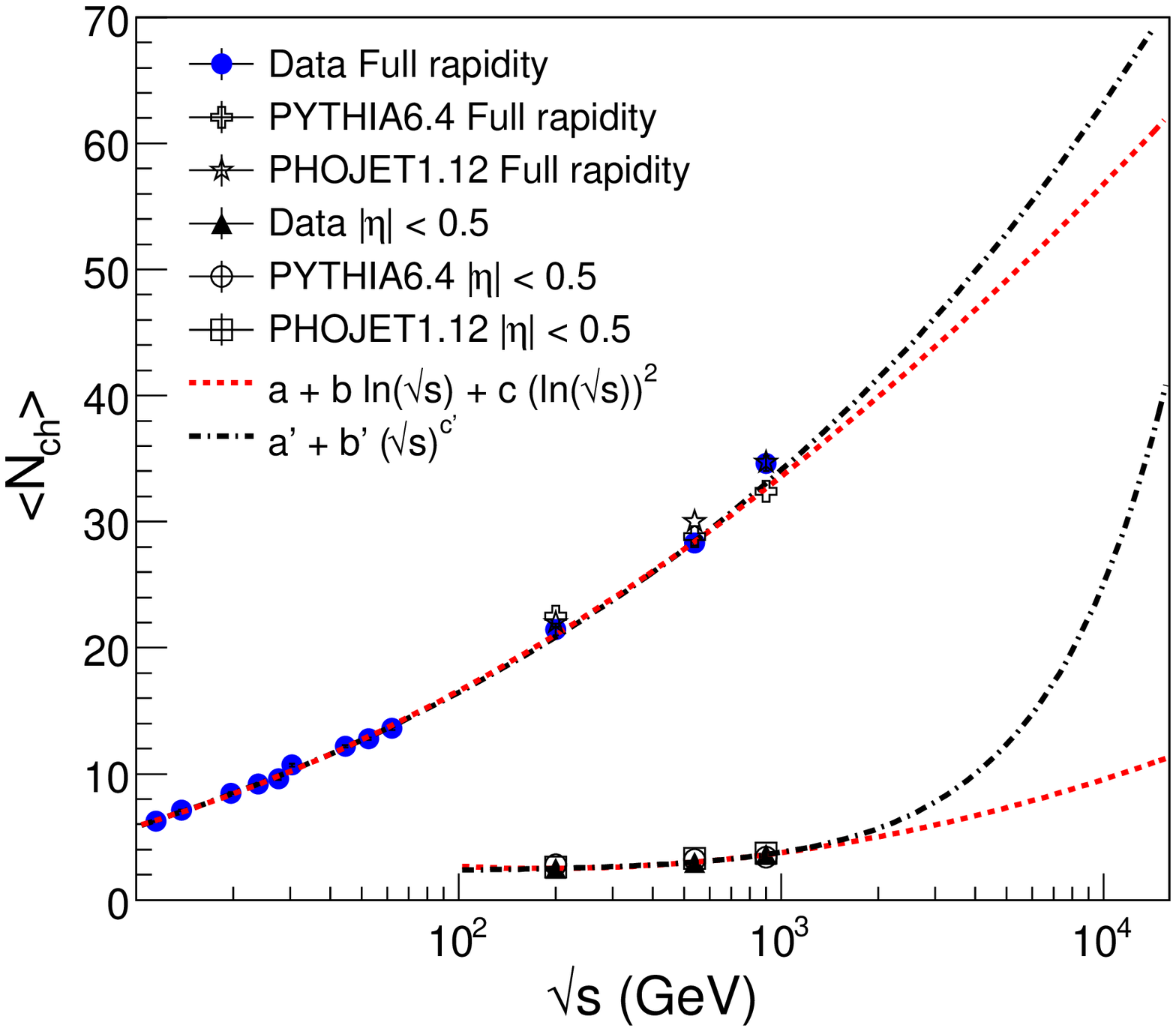}
 \includegraphics[scale=0.35]{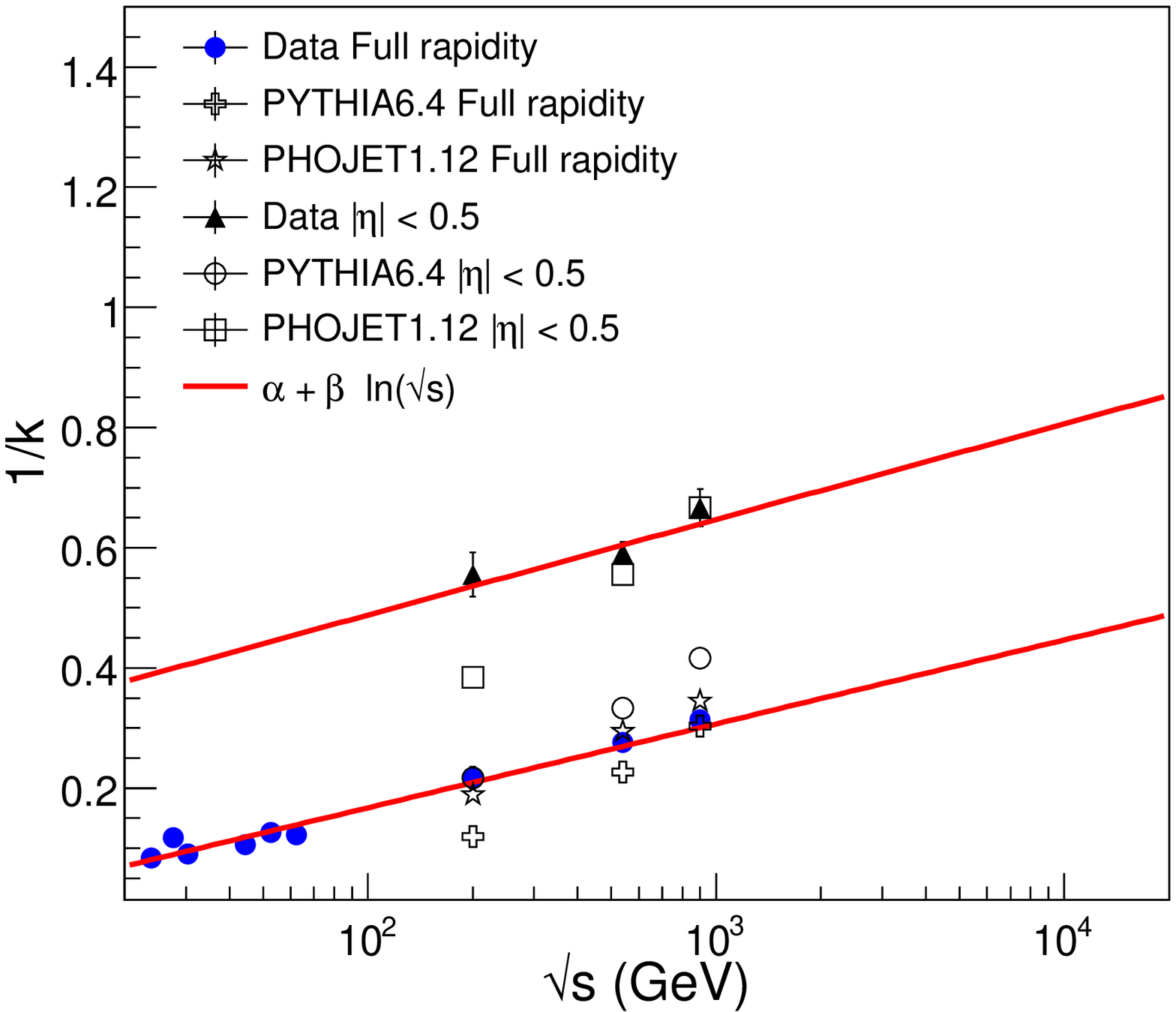}
\caption{NBD fit parameters $\langle N_{\mathrm {ch}} \rangle$ 
(left panel) and 1/$k$ (right panel)
to multiplicity distributions in $p$+$\bar{p}$ collisions at
various center of mass energies~\cite{datamult}.
The lines are fits to the data points using the function forms as discussed
in text. Also shown for comparison are the results from PYTHIA and PHOJET models
at $\sqrt{s}$ = 200, 540 and 900 GeV.}
\label{nbdparam}       
\end{center}
\end{figure}

The measurements of charged particle multiplicity distribution has been found to
be well described by negative binomial distribution (NBD) at midrapidity and also
for the full rapidity region in $p$+$\bar{p}$~\cite{datamult}. The NBD distribution has a form,
\begin{equation}
P_{\rm NBD}(\langle N_{\mathrm {ch}} \rangle, k; n) = \frac{\Gamma(n+k)}{\Gamma(n+1)\Gamma(k)} \cdot \frac{(\langle N_
{\mathrm {ch}} \rangle/k)^n}{(\langle N_{\mathrm {ch}} \rangle/k + 1)^{n+k}},
\label{nbdfit}
\end{equation}
The NBD has two parameters, $\langle N_{\mathrm {ch}} \rangle$ and $k$.
Where the parameter $k$ is an interesting quantity, 1/$k$ $\rightarrow$ 0 would
correspond to Poisson distribution (independent particle production) and $k$ = 1
would correspond to Geometric distribution. Under the limit of large multiplicity
($N_{ch}$ $\rightarrow$ Large), the NBD distribution goes over to a Gamma Distribution.
Some of the measured multiplicity distributions at midrapidity ($|\eta|<0.5$) and over full
pseudorapidity range ($|\eta|<5.0$) are fitted to NBD distribution and are
shown in Fig.~\ref{nbd} left and right panels respectively. 
The NBD parameters $\langle N_{\mathrm {ch}} \rangle$ and $k$
extracted are plotted in Fig.~\ref{nbdparam} left and right panels respectively.
The CDF experiment results are not used
in the current studies as they have multiplicity distributions with a high
$p_{T}$ cut-off of 400 MeV/$c$ measured within $|\eta|$ $<$ 1~\cite{cdf}.
PYTHIA and PHOJET model calculations suggest that there is more than 50\% loss
in the average number of charged particles due to a $p_{T}$ cut-off of 400 MeV/$c$
at midrapidity. Since all other results presented have a much smaller $p_{T}$ 
cut-off $\sim$ 100 MeV/$c$, inclusion of CDF results would make this comparative 
study heavily dependent on the model based extrapolation to lower $p_{T}$ regions.

\begin{figure}
\begin{center}
 \includegraphics[scale=0.35]{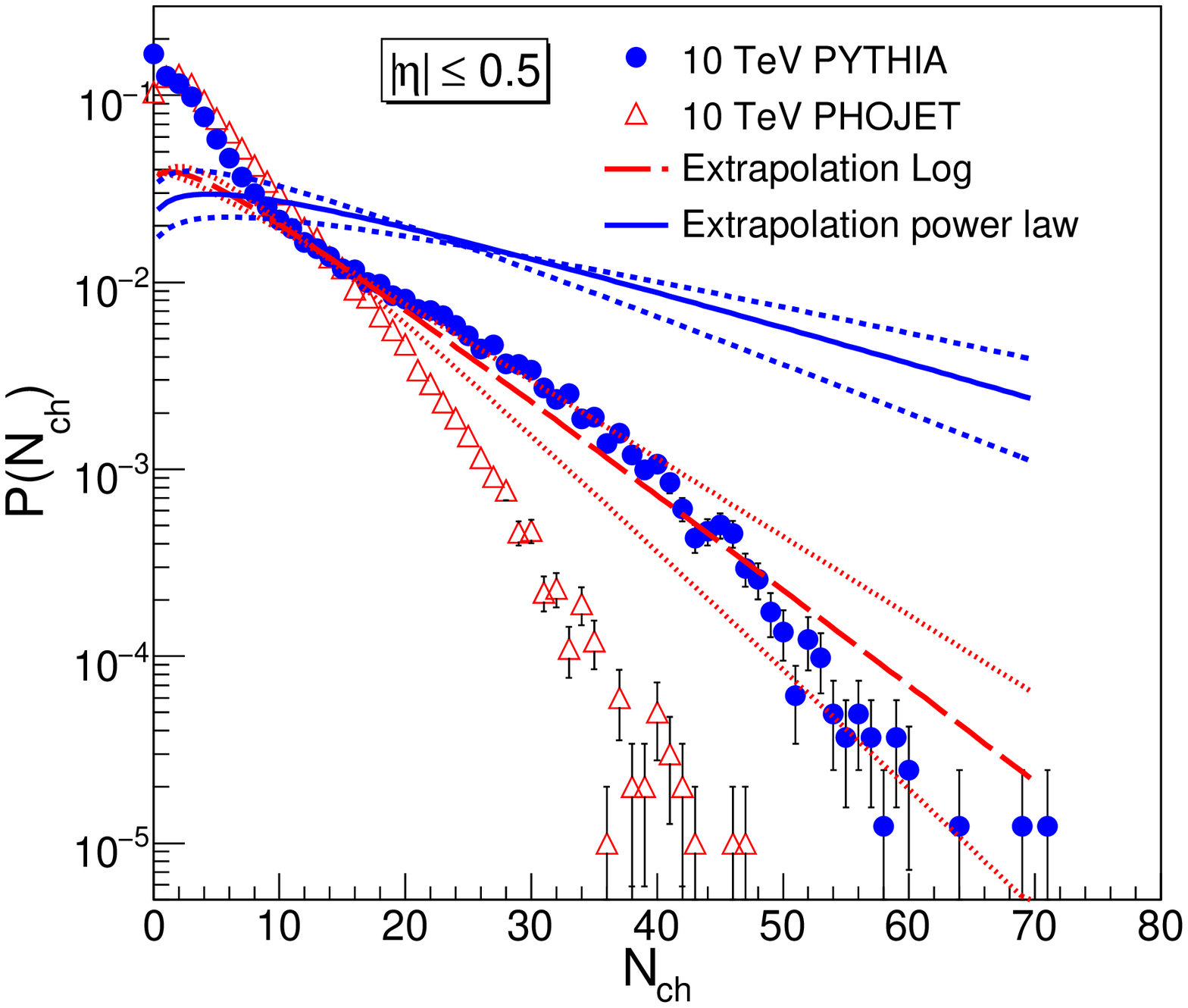}
 \includegraphics[scale=0.35]{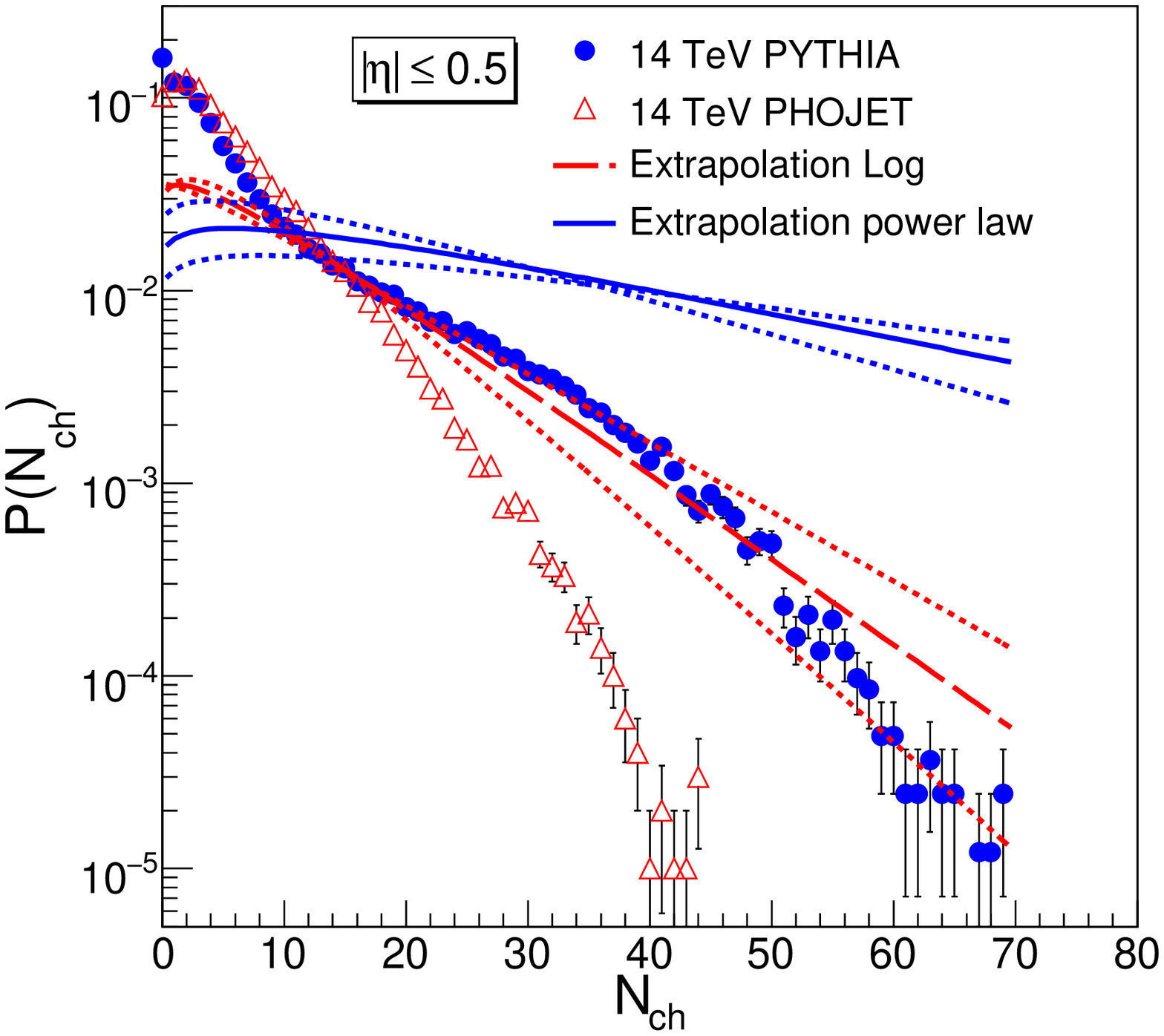}
\caption{Estimated multiplicity distribution for charged particles in
$p$+$p$ collisions at $\sqrt{s}$ = 10 (left panel) and 14 TeV (right panel) in midrapidity.
Solid and dashed lines are distributions obtained from $\langle N_{\mathrm {ch}} \rangle$
extrapolation using Eqns.~\ref{log} and ~\ref{power} respectively.
The dotted lines reflects errors in multiplicity distributions due to extrapolation
of the parameters $\langle N_{\mathrm {ch}} \rangle$ and $k$.
The results are compared to corresponding calculations from PYTHIA and PHOJET. }
\label{multpredictmid}       
\end{center}
\end{figure}

It is observed from Fig.~\ref{nbdparam} that both at midrapidity and full rapidity range
the $\langle N_{\mathrm {ch}} \rangle$ increases with $\sqrt{s}$ while
the $k$ value decreases with $\sqrt{s}$. The $\sqrt{s}$ region for which
measurements exists, the
$\langle N_{\mathrm {ch}} \rangle$ dependence can be described reasonably well
by both the following expressions (as shown in Fig.~\ref{nbdparam}),
\begin{equation}
\langle N_{\mathrm {ch}} \rangle = a~+~b~ln(\sqrt{s}) + c~(ln(\sqrt{s}))^{2}
\label{log}
\end{equation}
and
\begin{equation}
\langle N_{\mathrm {ch}} \rangle~=~a^{\prime}~+~b^{\prime}~(\sqrt{s})^{c^{\prime}},
\label{power}
\end{equation}
where $a$, $b$, $c$, $a^{\prime}$, $b^{\prime}$ and $c^{\prime}$ are fit parameters.
The values of $a$, $b$, $c$, $a^{\prime}$, $b^{\prime}$ and $c^{\prime}$ at midrapidity
are 3.8 $\pm$ 0.1, 1.5 $\pm$ 0.2, 0.45 $\pm$ 0.1, 2.3 $\pm$ 0.14, 1.5 $\pm$ 0.13
and 1.2 $\pm$ 0.26 respectively. For full rapidity coverage the values of 
$a$, $b$, $c$, $a^{\prime}$, $b^{\prime}$ and $c^{\prime}$ are 
1.3 $\pm$ 0.3, 0.62 $\pm$ 0.17, 0.59 $\pm$ 0.02, -10.5 $\pm$ 0.86, 9.9 $\pm$ 0.69
and 0.22 $\pm$ 0.01 respectively. It is noted that extrapolation of the power law 
function to LHC energies seems to lead to a sudden increase in average charged 
particle multiplicity at midrapidity. Such a large unexpected jump in the multiplicities 
at midrapidity already seems to put constraints on applicability of such a functional
form. As will be seen later in Fig.~\ref{multpredictmid} such a functional form will
lead to a probability distribution of charged particle multiplicity at midrapidity 
not showing the characteristic drop at higher multiplicity. Such a drop is expected
from PYTHIA, PHOJET models and is seen for available experimental data shown in Fig.~\ref{nbd}. 
In Fig.~\ref{nbdparam} the average number of charged particles from experiments at 
$\sqrt{s}$ = 200, 540 and 900 GeV are compared to corresponding results from PYTHIA and
PHOJET models. Both the models seem to be in reasonable agreement with the measurements
at midrapidity and full rapidity.

The 1/$k$ dependence on $\sqrt{s}$ has the form $\alpha$~+~$\beta$~$ln(\sqrt{s})$, where
$\alpha$ and $\beta$ are fit parameters. For midrapidity the values of $\alpha$ and $\beta$ 
are 0.65 $\pm$ 0.03 and 0.07 $\pm$ 0.03 respectively. For full rapidity case, the 
values of $\alpha$ and $\beta$ are -0.11  $\pm$ 0.006 and 0.06 $\pm$ 0.001 respectively.
The results from PYTHIA and PHOJET models are also shown. At higher energies the results 
from PHOJET is in better agreement with the measurements compared to those from 
PYTHIA. For the midrapidity measurements PHOJET model calculations
fail to match the data for $\sqrt{s}$ = 200 GeV. 

\begin{figure}
\begin{center}
 \includegraphics[scale=0.35]{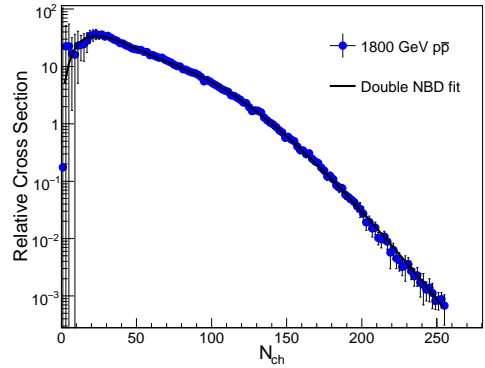}
\caption{Relative cross-section of charged particles produced in $p$+$\bar{p}$ collisions
at $\sqrt{s}$ = 1800 GeV~\cite{e735} fitted to a sum of to NBD functions. See text for more details.}
\label{e735}       
\end{center}
\end{figure}

We have examined the charged particle
multiplicity distribution from E735 Collaboration for $p$+$\bar{p}$ collisions at
$\sqrt{s}$ = 1800 GeV~\cite{e735}. The multiplicity distribution is found to be well explained
using the sum of two NBD functions instead of a single NBD as
for the rest of the data discussed above. The sum of two NBD is given as, 
\begin{equation}
F =  \omega P_{\rm NBD}^{1}(\langle N_{\mathrm {ch}} \rangle_{1}, k_{1}; n) + (1 - \omega) 
P_{\rm NBD}^{2}(\langle N_{\mathrm {ch}} \rangle_{2}, k_{2}; n)
\label{2nbdfit}
\end{equation}

where $P_{\rm NBD}^{1}$ and $P_{\rm NBD}^{2}$  have the same form as in Eqn~\ref{nbdfit},
$\omega$ is the weight factor, $\langle N_{\mathrm {ch}} \rangle_{1}$ and 
$\langle N_{\mathrm {ch}} \rangle_{2}$, $k_{1}$ and $k_{2}$ are the respective NBD 
parameters. The fit to the E735 data by this function is shown in Fig.~\ref{e735}.
The values of various parameters obtained from the fit are, 
$\langle N_{\mathrm {ch}} \rangle_{1}$ = 36.5 $\pm$ 1.7, 
$\langle N_{\mathrm {ch}} \rangle_{2}$ = 86.9 $\pm$ 2.6,
 $k_{1}$ = 2.8 $\pm$ 0.2,  $k_{2}$ = 10.3 $\pm$ 0.8 and 
$\omega$ = 0.18 $\pm$ 0.03. The overall average value of the
charged particle multiplicity is 44.4. Since this data could not be fitted
to a single NBD, in contrast to the data from other energies discussed in this paper,
we have not included this measurement in our extrapolation studies. Further 
it is observed from the Fig.~\ref{nbdparam} that the two extrapolations 
discussed in Eqns~\ref{log} and \ref{power}
tend to differ only above $\sqrt{s}$ = 2000 GeV. The 
physics reason attributed to a different shape of multiplicity distribution at $\sqrt{s}$ = 1800 GeV
compared to those at lower energies is due to multiple parton interactions~\cite{e735}.
These result are also indicative of the deviation from the KNO (Koba, Nielsen and Olesen)
scaling~\cite{kno}.
The weight factor shows the second NBD distribution dominates among the two. 
It will certainly be interesting to find out if this feature is more pronounced at the LHC energies.

\begin{figure}
\begin{center}
 \includegraphics[scale=0.35]{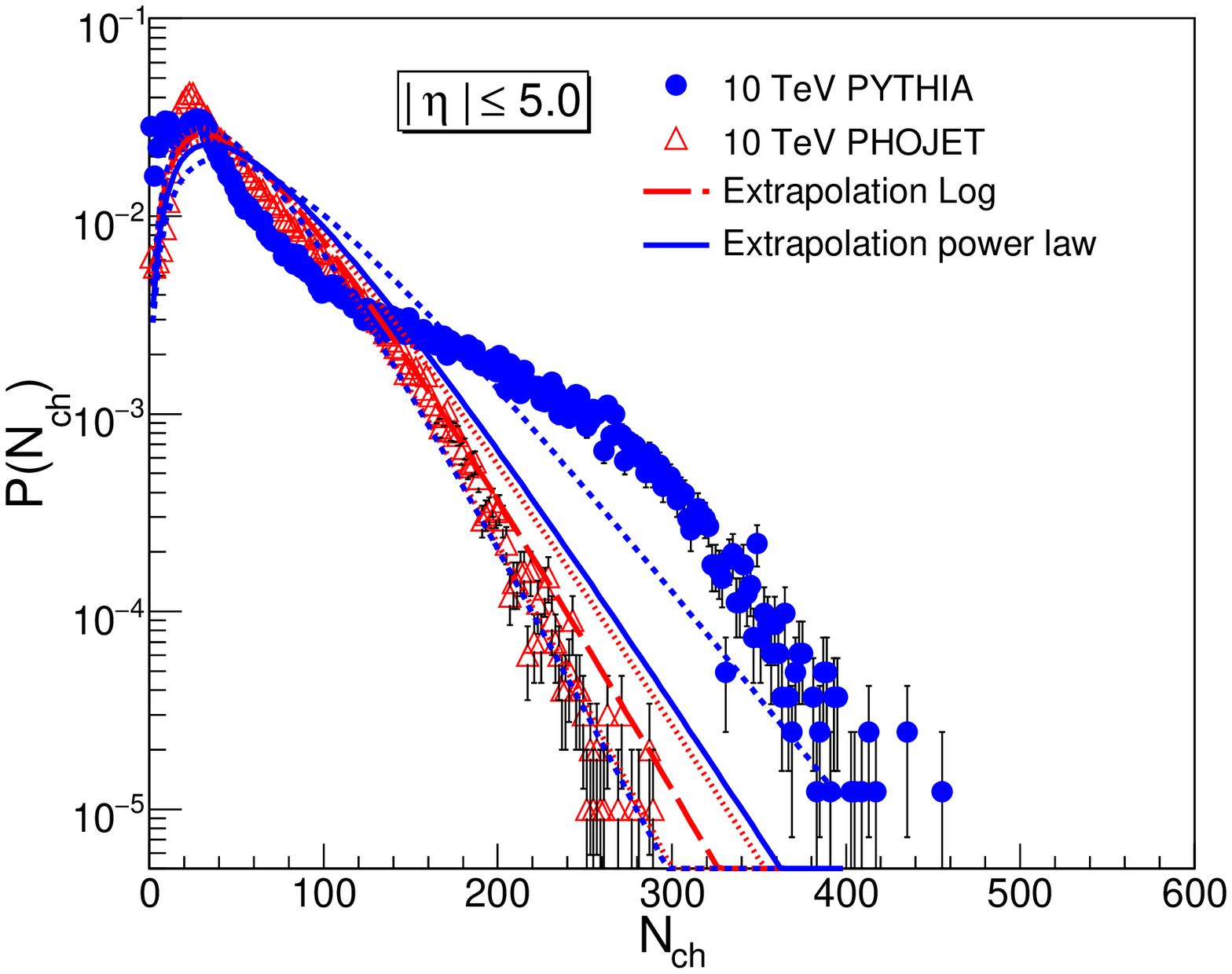}
 \includegraphics[scale=0.35]{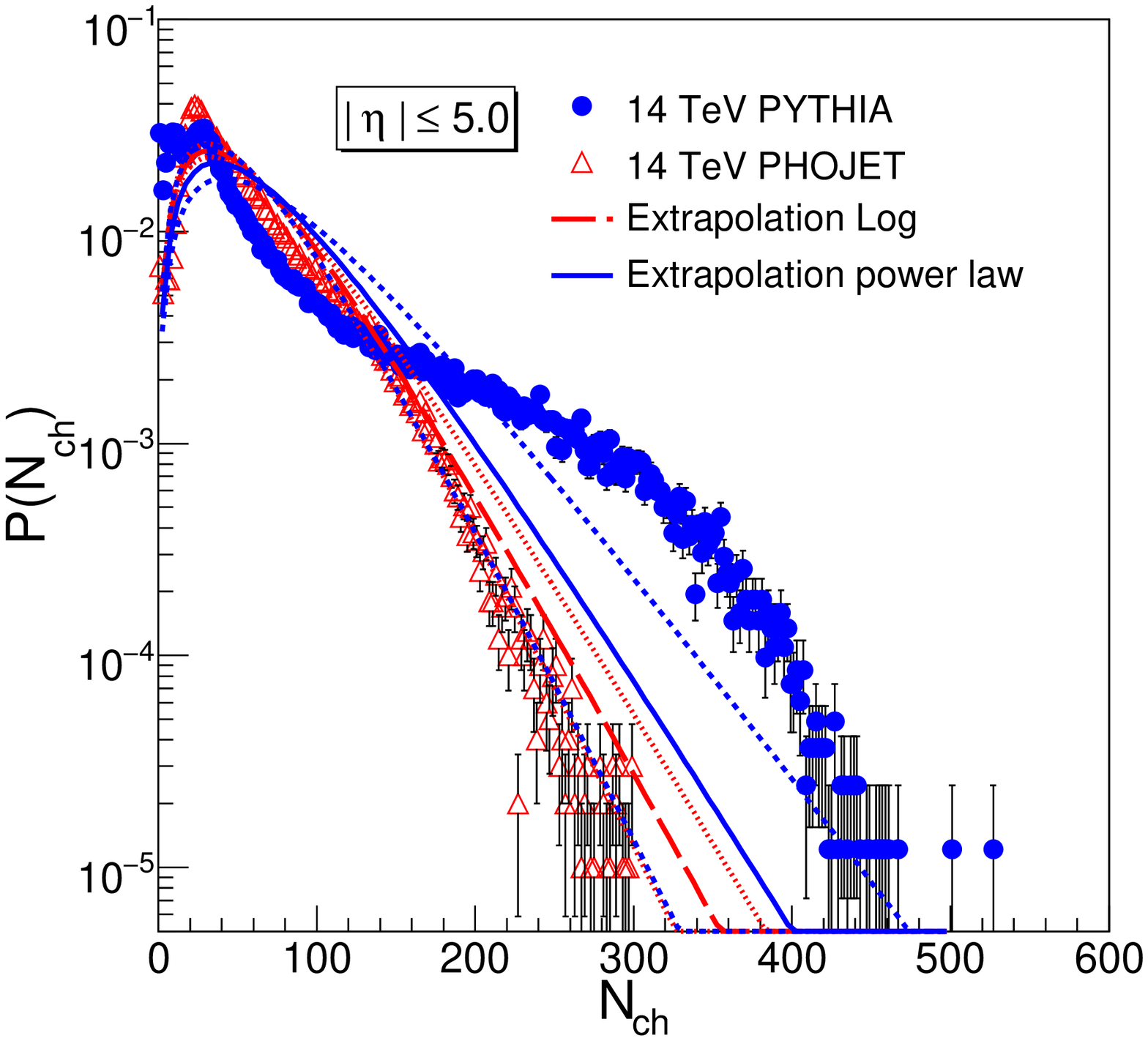}
\caption{Same as Fig.~\ref{multpredictmid} for full rapidity region.}
\label{multpredictfull}       
\end{center}
\end{figure}

The extrapolated values of  $\langle N_{\mathrm {ch}} \rangle$ and $k$ at both midrapidity and
full rapidity regions using the functional forms as in Eqns~\ref{log} and \ref{power} 
for $\sqrt{s}$ = 10 and 14 TeV are
given in the Table~\ref{nchktab}. Note the difference in values of $\langle N_{\mathrm {ch}} \rangle$
depending on the dependence on $\sqrt{s}$ as per Eqn.~\ref{log} (logarithmic dependence)
or Eqn.~\ref{power} (power law dependence). Knowing these values (parameters of NBD function)
we can predict the multiplicity distributions for both midrapidity and full rapidity ranges
at $\sqrt{s}$ = 10 and 14 TeV using Eqn.~\ref{nbdfit}. These distributions for both  midrapidity regions and
full rapidity regions are shown in Fig.~\ref{multpredictmid} and Fig.~\ref{multpredictfull}
respectively. The left panels are for $\sqrt{s}$ = 10 TeV and right panels are for 
$\sqrt{s}$ = 14 TeV. 
The results when compared to PYTHIA and PHOJET model calculations show that the
extrapolation of $\langle N_{\mathrm {ch}} \rangle$ using Eqn.~\ref{log} is favored
by the models at midrapidity. The extrapolated results show no such
preference to models in full rapidity region.
In general PYTHIA results are found to be higher compared to those from PHOJET
calculations. 
Actual experimental measurements at LHC will confirm the preferred $\langle N_{\mathrm {ch}} \rangle$ 
dependence on $\sqrt{s}$.

\begin{table}
\begin{center}
\caption{ Extrapolated NBD parameters $\langle N_{\mathrm {ch}} \rangle$ and $k$ for different $\sqrt{s}$
at midrapidity and full rapidity range for $p$+$p$ collisions at $\sqrt{s}$ = 10 and 14 TeV.
\label{nchktab}}
\begin{tabular}{ccccc}
\hline
Eqn.& $\sqrt{s}$ (TeV) & $\langle N_{\mathrm {ch}} \rangle$ & $k$ & $\eta$\
 range\\
\hline
Eqn.~\ref{log}      & 10 &  9.6  $\pm$ 1.1 & 1.2 $\pm$ 0.14  & $|\eta|<0.5$  \\
Eqn.~\ref{log}      & 10 &  56.7  $\pm$ 3.8 & 2.2 $\pm$ 0.09  & $|\eta|<5.0$   \\
Eqn.~\ref{power}    & 10 &  25.0  $\pm$ 11.5 & 1.24 $\pm$ 0.14  & $|\eta|<0.5$  \\
Eqn.~\ref{power}    & 10 &  63.7  $\pm$ 11.3 & 2.2 $\pm$ 0.09  & $|\eta|<5.0$   \\
Eqn.~\ref{log}      & 14 &  10.8  $\pm$ 1.3 & 1.20 $\pm$ 0.14  & $|\eta|<0.5$  \\
Eqn.~\ref{log}      & 14 &  60.7  $\pm$ 4.0 & 2.14 $\pm$ 0.09  & $|\eta|<5.0$  \\
Eqn.~\ref{power}    & 14 &  36.2  $\pm$ 18.5 & 1.20 $\pm$ 0.14  & $|\eta|<0.5$  \\
Eqn.~\ref{power}    & 14 &  69.4  $\pm$ 12.2 & 2.14 $\pm$ 0.09  & $|\eta|<5.0$  \\
\hline
\end{tabular}
\end{center}
\end{table}

\section{Pseudorapidity distribution}

\begin{figure}
\begin{center}
 \includegraphics[scale=0.35]{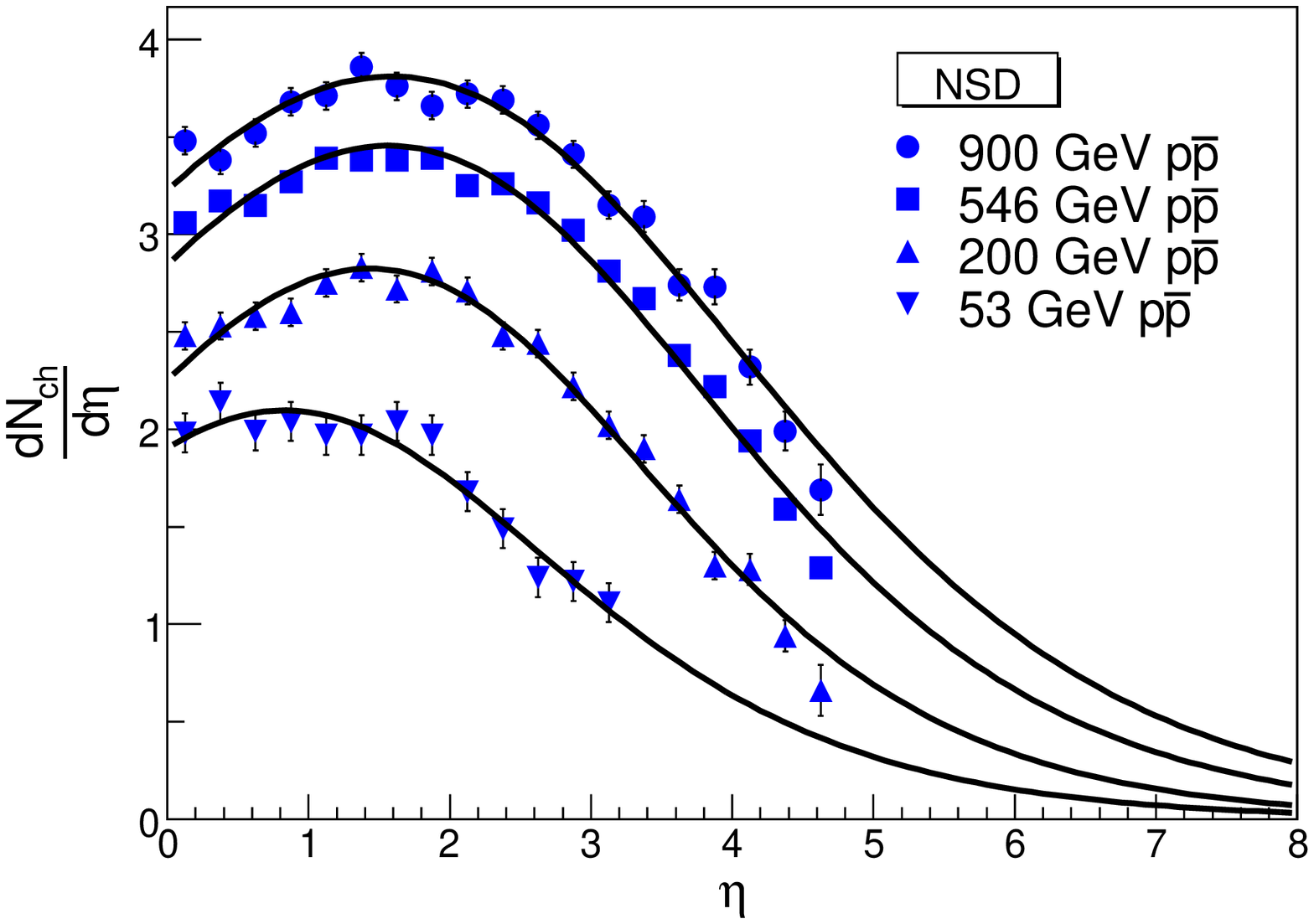}
 \includegraphics[scale=0.35]{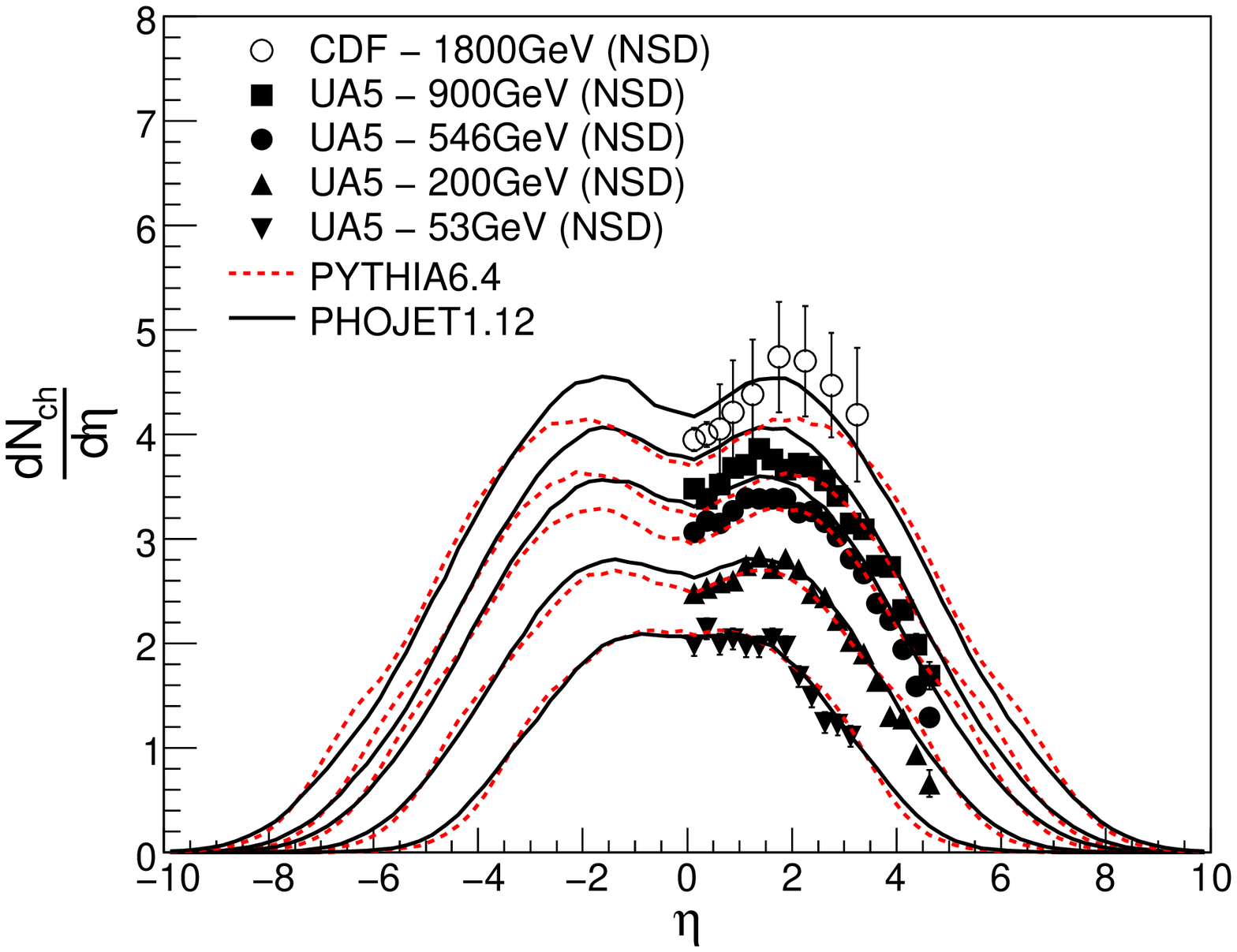}
\caption{Left panel: Pseudorapidity distribution for charged particles in
$p$+$\bar{p}$ collisions at various center of mass energies~\cite{datadnchdeta}.
The solid lines are fit to the data points using the function given in
Eqn.~\ref{dndeta}. Right panel: Comparison of the pseudorapidity distributions
for charged particles at various $\sqrt{s}$ to PYTHIA and PHOJET model calculations.}
\label{dnchdetafit}       
\end{center}
\end{figure}

The pseudorapidity distributions of charged particles from the existing
data at $\sqrt{s}$ = 53, 200, 546 and 900 GeV~\cite{datadnchdeta} can be described by the
following functional form,

\begin{equation}
\frac{dN}{d\eta} = \frac{C+\eta}{1 + {\exp} \frac{\eta - \eta_{0}}{\delta}}
\label{dndeta}
\end{equation}

This formula is chosen to describe the central plateau and the fall
off in the fragmentation region of the distribution by means of the
parameters $\eta_{0}$ and $\delta$ respectively. The term $C+\eta$,
$C$ is a fit parameter,
describes the magnitude of the distribution and the dip at the $\eta$ = 0.
A similar form has been used to describe the $p$+$p$ data at ISR energies~\cite{isr}
and heavy-ion collisions at RHIC~\cite{pmd}. The distribution is a generalization of
Fermi distribution and recent work suggests a relation of this functional form to 
string percolation model~\cite{spm}.
\begin{table}
\begin{center}
\caption{ Parameters $C$, $\eta_{0}$ and $\delta$ for different $\sqrt{s}$.
\label{dndytab}}
\begin{tabular}{ccccc}
\hline
Collision & $\sqrt{s}$ (GeV) & $C$ & $\eta_{0}$ & $\delta$\\
\hline
$p$+$\bar{p}$ & 53  &  2.4 $\pm$ 0.23 & 1.5 $\pm$ 0.17  & 1.12 $\pm$ 0.1  \\
$p$+$\bar{p}$ & 200 &  2.5 $\pm$ 0.07 & 2.5 $\pm$ 0.05  & 1.10 $\pm$ 0.04  \\
$p$+$\bar{p}$ & 546 &  3.0 $\pm$ 0.10 & 2.9 $\pm$ 0.07  & 1.15 $\pm$ 0.04  \\
$p$+$\bar{p}$ & 900 &  3.6 $\pm$ 0.10 &  3.0 $\pm$ 0.05  & 1.36 $\pm$ 0.05  \\
\hline
\end{tabular}
\end{center}
\end{table}

The values of the parameters $C$, $\eta_{0}$ and $\delta$ obtained by fitting the
data distributions with Eqn.~\ref{dndeta} are given in Table~\ref{dndytab}
and the fits to data are shown in Fig.~\ref{dnchdetafit} (left panel).
The value of parameters $C$ and $\eta_{0}$ are found to increase
with increasing $\sqrt{s}$. The value of the parameter
$\delta$ is found to be approximately independent of $\sqrt{s}$ within errors.
The constancy of $\delta$ is another way of demonstrating the concept of limiting
fragmentation~\cite{lf}. In such a scenario, multiplicity density in pseudorapidity
when plotted as a function of pseudorapidity shifted by beam rapidity is
expected to be independent of pseudorapidity at forward rapidities~\cite{pmd}.
Shown in the right panel of Fig.~\ref{dnchdetafit} are the comparisons of the
experimentally measured charged particle pseudorapidity distributions to those
from the PYTHIA and the PHOJET calculations. The CDF measurements~\cite{cdf2} are not used to 
obtain the parameters $C$, $\eta_{0}$ and $\delta$ for making predictions at
LHC energies as discussed later, because of their very limited $\eta$ coverage.
It is observed that for the lowest beam energy studied, $\sqrt{s}$ = 53 GeV, both PYTHIA and PHOJET are in
very good agreement with the experimental data. As the beam energy increases 
PYTHIA results seems to be in better agreement with the data. While for the 
top energy ($\sqrt{s}$ = 1800 GeV~\cite{cdf}) studied the error bars 
are large to make a conclusion on which of the two models have better agreement
with the data.

\begin{figure}
\begin{center}
\includegraphics[scale=0.35]{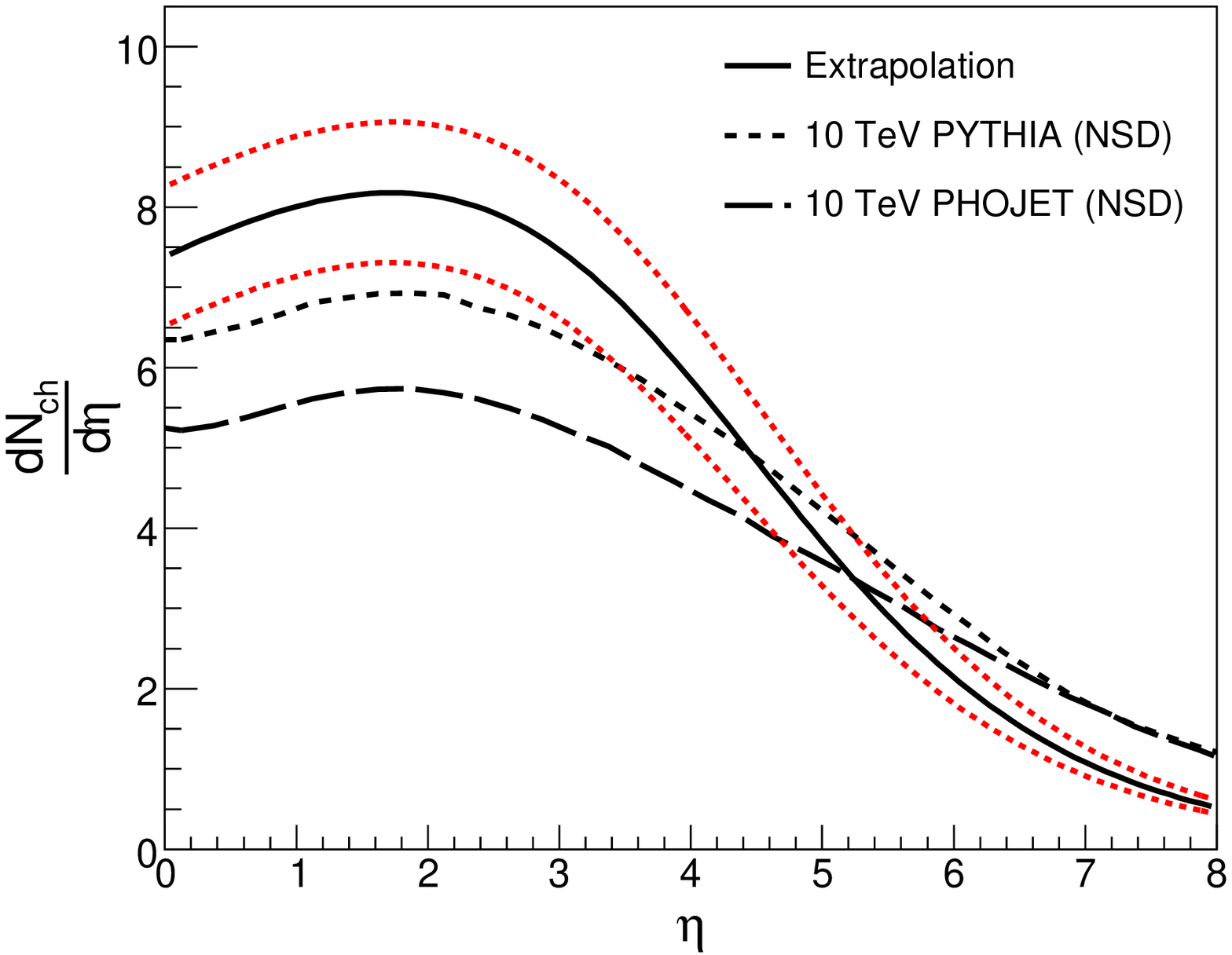}
 \includegraphics[scale=0.35]{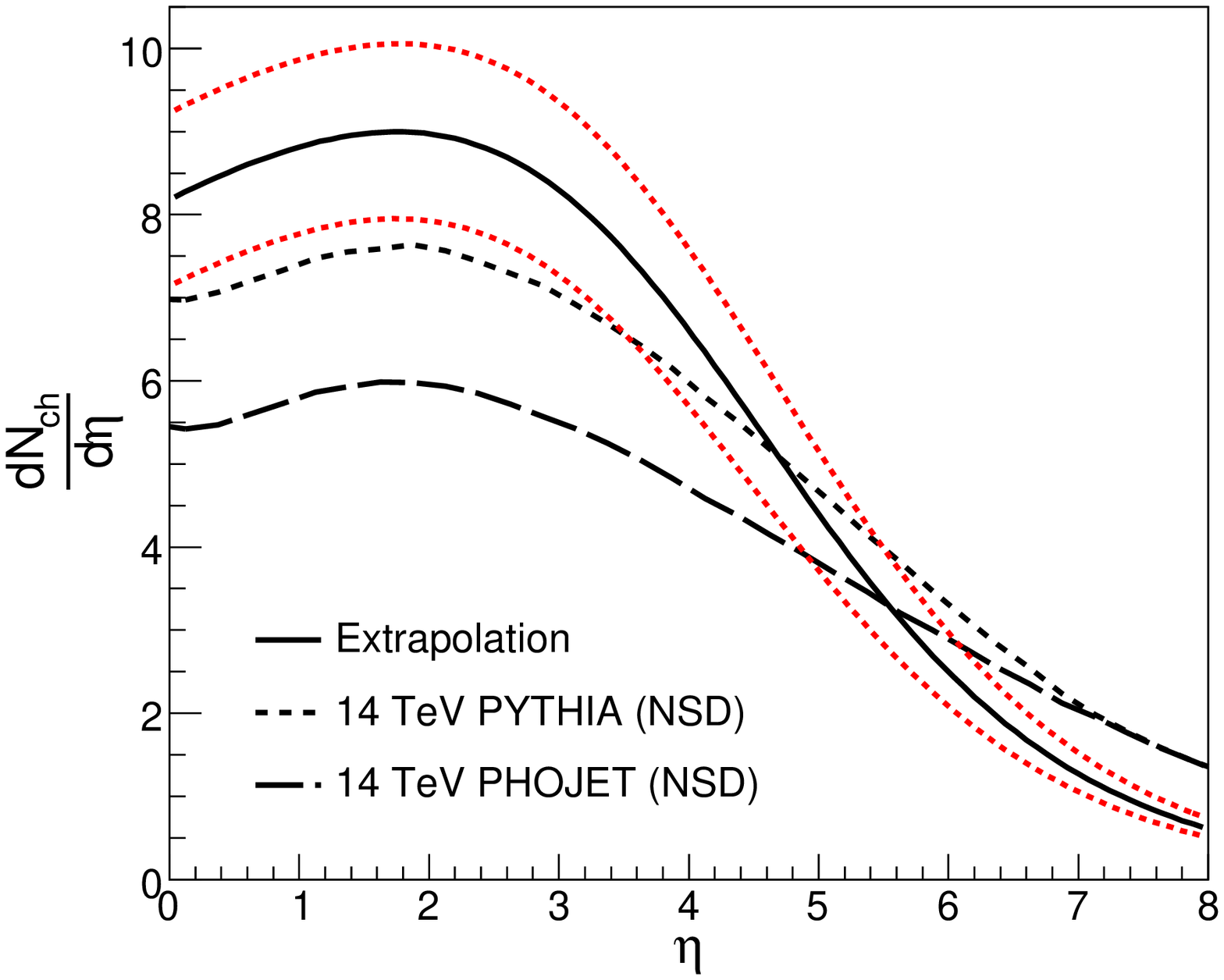}
\caption{Expected pseudorapidity distribution (solid lines) for charged particles in
$p$+$p$ collisions at $\sqrt{s}$ = 10 (left panel) and 14 TeV (right panel). 
This is obtained from the extrapolation
using the existing data from $p$+$\bar{p}$ collisions at lower energies. The dotted
lines indicate the uncertainties associated with the extrapolation. Also shown
for comparison are the expected $dN_{\mathrm {ch}}/d\eta$ from PYTHIA and PHOJET
model calculations for $p$+$p$ collisions at $\sqrt{s}$ = 10 and 14 TeV.}
\label{predicdndy}       
\end{center}
\end{figure}

Using the average value of $\delta$ and extrapolating the value of $C$ and $\eta_{0}$
to $\sqrt{s}$ = 10 and 14 TeV, we are able to predict the full pseudorapidity distribution
for charged particles. The extrapolation is done by fitting the variation of $C$
with $\sqrt{s}$ with a functional form

\begin{equation}
C~=~3.7~+~1.15~ln(\sqrt{s})~+~0.25~(ln(\sqrt{s}))^{2}
\end{equation}
and the variation of $\eta_{0}$ with $\sqrt{s}$ as
\begin{equation}
\eta_0~=~3.1~+~0.4~ln(\sqrt{s}).
\end{equation}
The values of $C$ obtained are 7.63 $\pm$ 0.87 and 8.43 $\pm$ 1.04 for
$\sqrt{s}$ = 10 and 14 TeV respectively. Those for $\eta_{0}$ are 4.01 $\pm$ 0.14
and 4.15 $\pm$ 0.15 for $\sqrt{s}$ = 10 and 14 TeV respectively.
Assuming the functional form given in Eqn.~\ref{dndeta} is valid for
$\sqrt{s}$ = 10 and 14 TeV and using the parameter values obtained as
above, we can now predict the full pseudorapidity distribution of charged particles
at $\sqrt{s}$ = 10 and 14 TeV. These distributions are shown in Fig.~\ref{predicdndy}
along with expectations calculated from PYTHIA and PHOJET models for the same collisions.
The left panel shows the results for $\sqrt{s}$ = 10 TeV and right panel shows the
results for $\sqrt{s}$ = 14 TeV.

It is observed that at both $\sqrt{s}$ = 10 and 14 TeV, the predictions from
a judicious extrapolation of existing data in general are above the model predictions.
The PYTHIA results are close to the lower error band (dotted lines) for $\eta$ $<$ 4.
The values from PYTHIA are higher than those from PHOJET model calculation
over a large pseudorapidity range presented. In general all the three distributions
have almost similar shape. The differences between the models could arise due to several reasons.
The event generation in PYTHIA is mainly designed to describe the 
possible hard interactions in $p$+$p$($\bar{p}$) collisions. It also
combines sophisticated models dealing with soft hadronic
interactions~\cite{pythia}. However in the case of PHOJET, 
the main approach is to describe the soft component
of hadron-hadron, photon-hadron interactions at high energies. 
It combines the hard component calculated by perturbative QCD 
at the partonic level~\cite{phojet}.
Due to the different underlying theoretical models used
PYTHIA has larger model parameters which can be adjusted in order 
to better reproduce the data compared to PHOJET. Another difference lies in the 
parametrisation used get the $p$+$p$($\bar{p}$) cross-sections.
PYTHIA uses those derived from the Pomeron
exchange model while the PHOJET uses the optical theorem 
and cross-sections are corrected for high energies using 
the unitarity principle.

\section{Summary}
We have obtained the $\langle N_{\mathrm {ch}} \rangle$, $N_{\mathrm {ch}}$ distribution
and $dN_{\mathrm {ch}}/d\eta$ versus $\eta$ for $\sqrt{s}$ = 10 and 14 TeV using extrapolations
of existing data. The results have been compared to calculations from PYTHIA and PHOJET.
Measurements at midrapidity at LHC will help distinguish whether $\langle N_{\mathrm {ch}} \rangle$
has a power law or a logarithmic dependence on $\sqrt{s}$. The shape of $dN_{\mathrm {ch}}/d\eta$ versus $\eta$ obtained from the extrapolation of information from existing measurements are very similar to
those calculated from models. PYTHIA calculations of $N_{\mathrm {ch}}$ and
 $dN_{\mathrm {ch}}/d\eta$ distributions seem to be higher compared to those from PHOJET model.
The $N_{\mathrm {ch}}$ distributions from models suggest the multiplicity distribution at LHC
energies may not be well described by a single NBD distribution, already such a behaviour
has been seen at $\sqrt{s}$ = 1.8 TeV~\cite{e735}.\\

\noindent{\bf Acknowledgments}\\
Financial assistance from the Department of Atomic Energy, Government 
of India is gratefully acknowledged.\\

%
%

\end{document}